\documentclass[conference]{IEEEtran}
\IEEEoverridecommandlockouts
\usepackage{amsmath,amssymb,amsfonts}
\usepackage{algorithmic}
\usepackage{graphicx}
\usepackage{textcomp}
\usepackage{xcolor}
\usepackage{diagbox}
\usepackage[
backend=biber,
style=ieee,
sorting=none
]{biblatex}
\def\BibTeX{{\rm B\kern-.05em{\sc i\kern-.025em b}\kern-.08em
    T\kern-.1667em\lower.7ex\hbox{E}\kern-.125emX}}
\begin{document}

\title{Deep Learning-based RF Fingerprint Authentication with Chaotic Antenna Arrays\\
\thanks{This work was funded by U.S. National Science Foundation (NSF) under grant 2233774.}
}

\author{\IEEEauthorblockN{1\textsuperscript{st} Justin McMillen}
\IEEEauthorblockA{\textit{Dept. of Electrical Engineering} \\
\textit{University of South Florida}\\
Tampa, United States \\
jmcmillen@usf.edu}
\and
\IEEEauthorblockN{2\textsuperscript{nd} Gokhan Mumcu}
\IEEEauthorblockA{\textit{Dept. of Electrical Engineering} \\
\textit{University of South Florida}\\
Tampa, United States \\
mumcu@usf.edu}
\and
\IEEEauthorblockN{3\textsuperscript{rd} Yasin Yilmaz}
\IEEEauthorblockA{\textit{Dept. of Electrical Engineering} \\
\textit{University of South Florida}\\
Tampa, United States \\
yasiny@usf.edu}}

\maketitle

\begin{abstract}
Radio frequency (RF) fingerprinting is a tool which allows for authentication by utilizing distinct and  random distortions in a received signal based on characteristics of the transmitter. We introduce a deep learning-based authentication method for a novel RF fingerprinting system called Physically Unclonable Wireless Systems (PUWS). An element of PUWS is based on the concept of Chaotic Antenna Arrays (CAAs) that can be cost effectively manufactured by utilizing mask-free laser-enhanced direct print additive manufacturing (LE-DPAM). In our experiments, using simulation data of 300 CAAs each exhibiting 4 antenna elements, we test 3 different convolutional neural network (CNN) architectures under different channel conditions and compare their authentication performance to the current state-of-the-art RF fingerprinting authentication methods.
\end{abstract}


\section{Introduction}

As the number of Internet of Things (IoT) devices and amount of wireless data communication rapidly increase, so does the threat posed by adversarial parties trying to exploit the vulnerabilities of wireless systems. Hence, it is vital to develop more secure methods of authentication and communication while satisfying the quality and efficiency constraints. With current technology, security at higher levels in the system (software) usually cannot protect lower layers (such as spoofing and jamming of hardware). Therefore, to work towards a more secure system, physical layer security features can and must complement the upper layer defenses, e.g., multi-factor authentication through RF fingerprinting. With ever increasing technology available to attackers and the emergence of much faster computing methods, traditional encryption techniques will not always be as secure as they currently are \cite{365700}. With this in mind, it is imperative to find new, harder-to-crack hardware-based methods to allow for secure systems.

RF fingerprinting is a promising authentication technique for physical layer security. The classical RF fingerprinting methods utilize the small amplitude, phase, and frequency variations that are unique to each device due to the inevitable randomness during the fabrication of the RF integrated circuits (ICs) connected to the antenna elements \cite{4261134,5601959}. These signatures, while being detectable by Machine Learning (ML) algorithms \cite{3}, are extremely small due to tight IC fabrication tolerances. State-of-the-art deep neural networks can only achieve around 63\% accuracy in authenticating 250 devices using these small signatures \cite{9063411}.   

We investigate the concept of \emph{Physically Unclonable Wireless Systems} (PUWS) as a new hardware security architecture to augment RF fingerprinting-based authentication. PUWS can effectively be built by additive manufacturing approaches where multiple devices within the structure can be randomized to exhibit enhanced and distinct fingerprints. An important element for PUWS is chaotic antenna arrays (CAAs) where antenna shapes, locations, and feed networks are randomized \cite{karabacak2021arraymetrics}. The prior work \cite{karabacak2021arraymetrics} on CAA assumed that the user with CAA has knowledge of the wireless channel, its own phase errors, and with no spatially varying phase errors. Such knowledge by the device is undesired and mostly impractical for real life applications. In this paper, we extend the CAA concept to work \emph{without} the knowledge of wireless channel or its own errors during authentication by resorting to deep learning-based detection algorithms. The mask-free laser-enhanced direct print additive manufacturing (LE-DPAM) technology investigated by our group can play as the key enabler of such antenna arrays\cite{8888323,8072664,9276426,8497031}. Introducing randomness will generate RF fingerprints based on phase errors at the antenna elements with spatial (i.e., $\theta$, $\phi$) variance, which will be shown to greatly benefit physical layer authentication with RF fingerprint enhancement.

\section{Chaotic Antenna Array Model}
\begin{figure}
    \centering
    \includegraphics[width=0.3\textwidth]{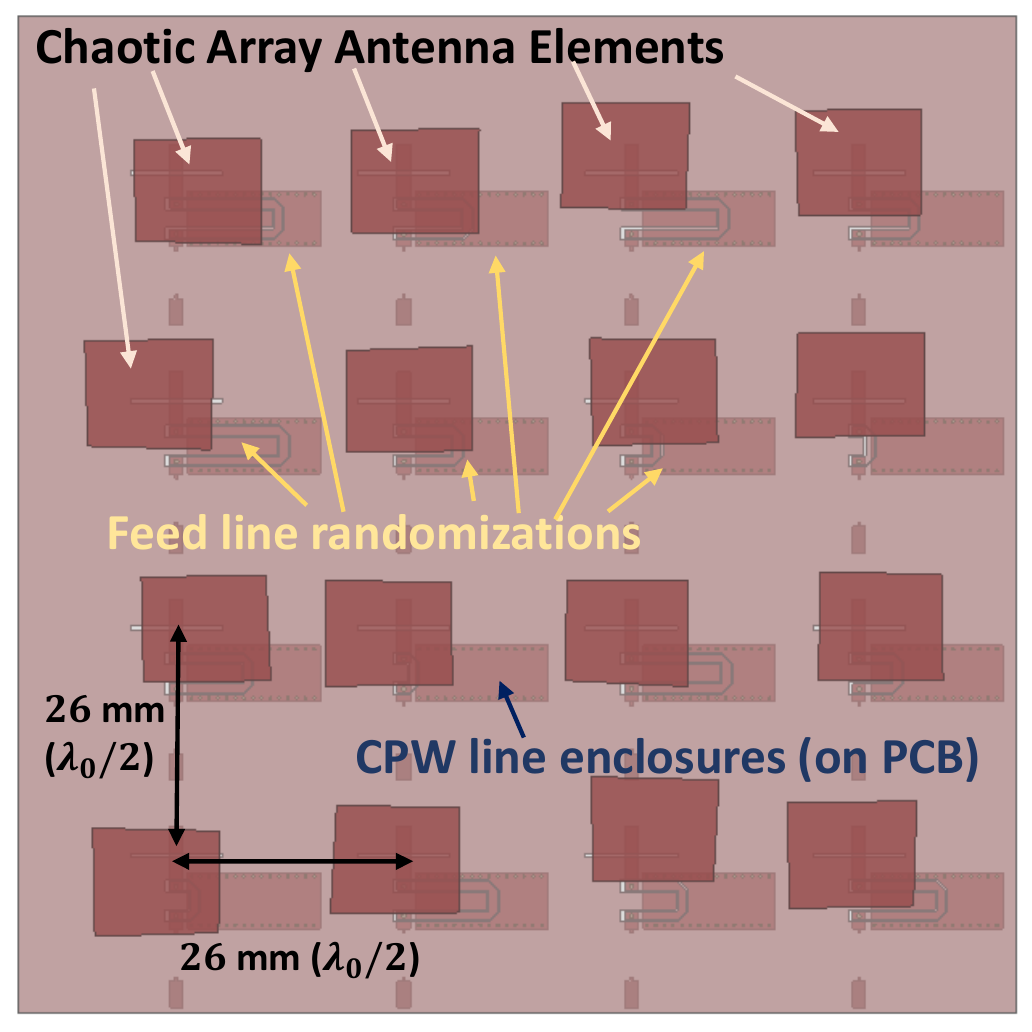}
    \caption{Ansys model of a 4$\times$4 5.8 GHz ISM band CAA with randomized antenna positions, shapes, and feed line lengths.}
    \label{fig:caa}
\end{figure}

For the model, we consider aperture coupled rectangular patch antennas and introduce randomness into the geometry of the patch and the length of a meandered feed line section inserted between the coupling aperture and feed point. The superstrate, substrate, and coupling aperture thicknesses are kept constant. The unperturbed patch is 14.4 x 12 mm$^2$ and resonates at 5.8 GHz with 12\% matching bandwidth, well above ISM bandwidth, allowing resonance shifts by these randomizations to be tolerable. Corner points $(x_i,y_i), \,i \,\in\, \{1,2,3,4\}$ are randomized according to:
\begin{equation}
    (x_i,y_i) = (x_i = r\cos\psi + c_i\cos\tau_i, y_i + r\sin\psi + c_i\sin\tau_i)
\end{equation}
where $r \,\in\, U[0,R]$ and $\psi \,\in\, U[0,2\pi]$, denoting a randomized shift in the center point of the rectangle with respect to the coupling aperture. $c_i \,\in\, U[0,C]$ and $\tau_i \,\in\, U[0,2\pi]$ are used to perturb the corners of the shifted rectangle to change its shape into a trapezoid. $U$ represents a uniform distribution while $R$ and $C$ set maximum limits for the randomizations, set to 4mm and 0.5mm, respectively. Figure \ref{fig:caa} shows a 16 element CAA that can be formed from the randomized antennas. The CAA technique is suitable for any frequency band and 5.8 GHz ISM is selected for popularity and availability of components to form a test bed in near future. Antenna spacing is half-wavelengths as in traditional practice.

\begin{figure}
     \centering     \includegraphics[width=0.5\textwidth]{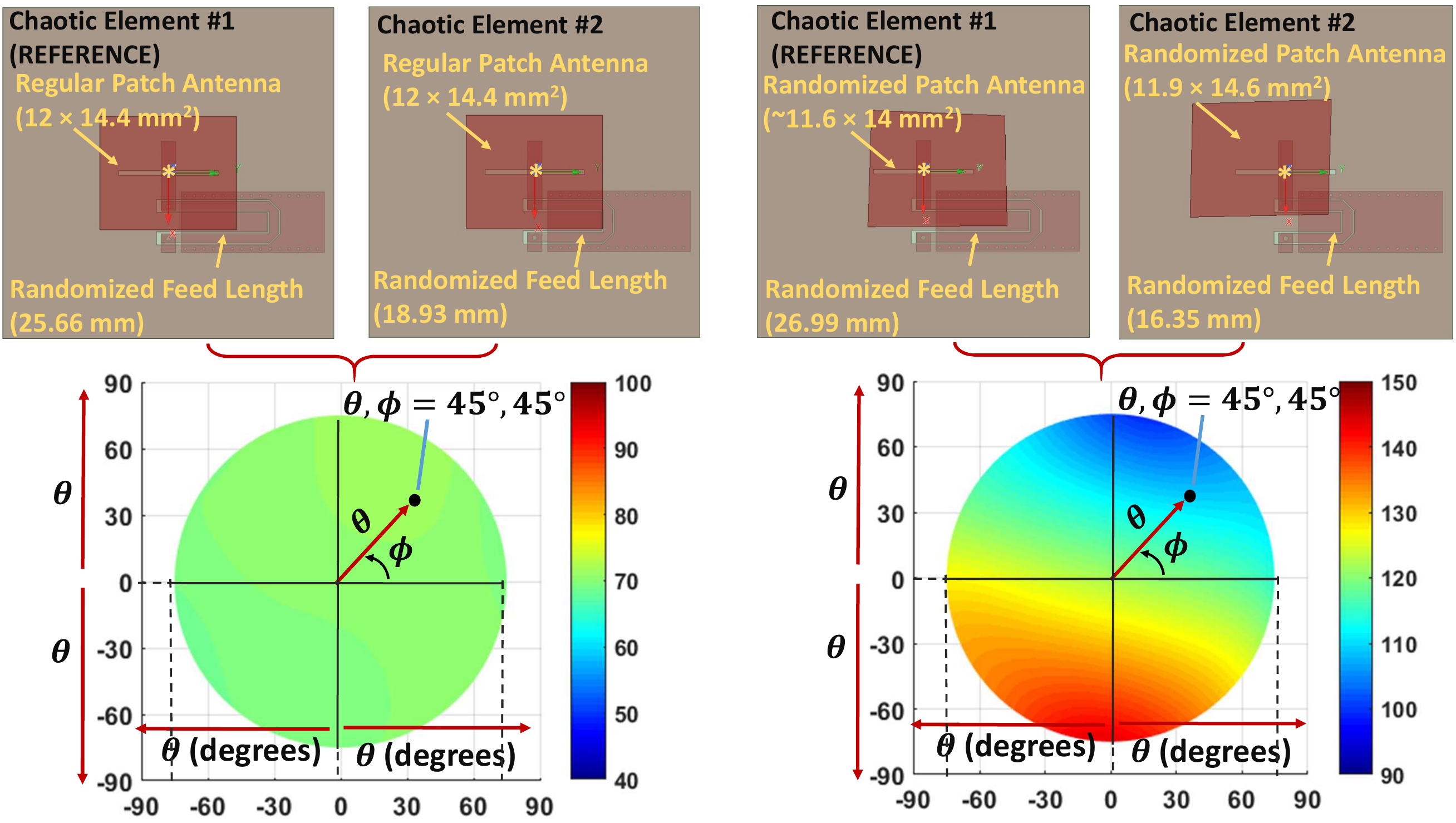}
     \caption{Phase error w.r.t. a reference antenna in polar coordinates. (Left) Feed line randomization generates a constant error in all transmission directions; (Right) Antenna geometry randomization creates $\theta,\phi$ dependent error.}
     \vspace{-5mm}
\label{fig:ElementCompare}
\end{figure}

Randomizing the antenna shape provides the spatial variance property depicted in Fig. \ref{fig:ElementCompare}. While feed line length randomization alone creates an enhanced signature transmitted equally in all directions, like the traditional RF fingerprint, antenna shape randomization causes the phase error to depend on the direction of radiation as evidenced by the colorful phase distribution in Fig.\ref{fig:ElementCompare}. Histogram  data obtained from 1200 antennas shows that phase error is uniformly distributed across a ~[0, $\sim2\pi$] range.
 
\section{Authentication}
The present state-of-the-art in RF fingerprinting-based authentication relies on deep neural networks \cite{9063411,8737463,9277909}, as opposed to earlier schemes that used traditional ML techniques (e.g., SVM, kNN, etc.) \cite{danev2009physical,ZHUO2017472}, statistical detectors \cite{7208645}, and wavelet transforms. Modern deep convolutional neural networks (CNNs) can successfully authenticate naturally occurring signatures in the RF chain in idealized setups with a small number of devices, according to recent literature \cite{8737463,9277909}. However, \cite{9063411} recently showed in a sizable study that even for cutting-edge deep CNNs, \emph{naturally occurring RF fingerprints are insufficient under realistic circumstances with a large number of devices and changing channel conditions between training and testing}.

\begin{figure}
     \centering     \includegraphics[width=0.5\textwidth]{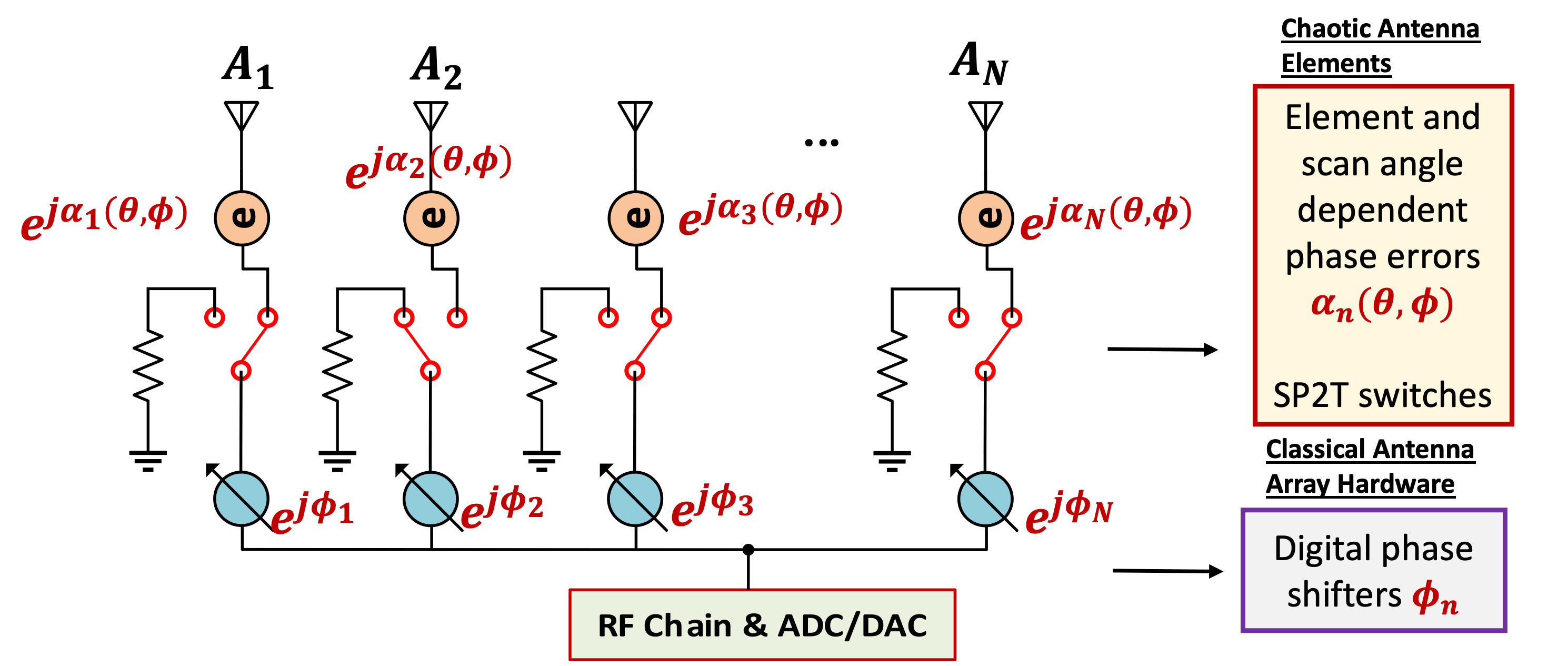}
     \caption{Circuit diagram of CAA. Each chaotic antenna element is sequentially turned on using switches and has a random and direction-dependent phase error due to its unique geometry.}
     \vspace{-5mm}
\label{fig:circuit}
\end{figure}

\subsection{Simulation Setup}
For a study on the feasibility of RF fingerprint authentication on PUWS, we used simulation data describing the phase variation of 1200 antenna elements and formed 300 CAAs each with 4 antenna elements. The elevation angle $\theta$ and the azimuth angle $\phi$ from the transmitting CAA to the receiver is randomly selected within [0\textdegree, 75\textdegree] and [-180\textdegree, 180\textdegree], respectively. We simulated a $f=5$ GHz WiFi environment with Rayleigh multipath fading, in which people may be walking between the device and the router. Considering a walking speed $V_{walk}$ %ranging from 1 m/sec to 10 m/sec, 
of 1 m/sec, the maximum Doppler shift $f_d$ is %between 16.67 Hz and 166.7 Hz 
calculated as 16.67 Hz
using the formula $f_d=(V_{walk}/c)f$, with the channel coherence time under the Clarke’s model, $T_c=\sqrt{9/{16\pi}f_d^2}=0.0254$ sec. %, ranging from approximately $0.0254$ to $0.00254$ sec. %Also, the sample rate $F_s$ was varied between from 10 KHz to 1 MHz. 
With a sampling rate of 1 MHz, $N = 1000$ samples are collected within an authentication sequence of 1 msec.
%Collecting $N = 1,000$ samples within an authentication sequence we obtained datasets for different scenarios. 

In each authentication sequence, the 4 antennas in a CAA are turned on sequentially to transmit a complex pilot signal, as illustrated in Fig. \ref{fig:circuit}. The circuit diagram in Fig. \ref{fig:circuit} also shows the chaotic antenna elements with random and direction-dependent phase errors together with the classical digital phase shifters. The authenticator receives the in-phase and quadrature (I/Q) samples through Rayleigh multipath fading channels in addition to additive white Gaussian noise: 
\begin{equation}
    y_n = h_n x_n + w_n,
\end{equation}
where $x_n=e^{j\alpha_n}$ is the transmitted pilot signal with constant amplitude and the corresponding phase error $\alpha_n$, $h_n\sim \mathcal{N}_c(0,\sigma_h^2)$ is the zero-mean complex Gaussian channel coefficient, and $w_n\sim \mathcal{N}_c(0,\sigma_w^2)$ is the additive white complex Gaussian noise. The I and Q samples are the real and imaginary parts of the received signal. With 4 antenna elements in each array and collecting I and Q samples of the received signal from each antenna, the dataset used to authenticate a CAA has a size of $N \times 8$, where $N=1000$ is the number of samples within an authentication session. The CAA phase errors and the received signals through Rayleigh fading channels were simulated in Matlab. 
%Power amplifier non-linearities were applied to each CAA using a Volterra series, $f_{PA}(x_t) = x_t(1+\psi_0|x_t|^2+\psi_1|x_t|^4)$, where $x_t$ is the most recent I/Q sample, and $\psi_0$, $\psi_1$ are coefficients unique to each power amplifier. 
The signal-to-noise ratio (SNR) is set to 20 dB in the simulations, i.e., $\sigma_w^2=0.01$. The CAA data is separated using a (80-10-10)\% split for training, validation, and testing, respectively. %By varying $V_{walk}$ %and $F_s$, 7 
%3 different datasets were created to allow testing under different scenarios. Table \ref{table:dataset} summarizes the different datasets and their properties. 

%\begin{table}
%    \centering
%    \caption{Dataset Parameters}
%    \begin{tabular}
%        {|c|| c | c | c |} %c | c | c | c |} 
%        \hline
%        \backslashbox{Variable}{Dataset} & 1 & 2 & 3 \\ [0.5ex] % & 4 & 5 & 6 & 7 \\ [0.5ex] 
%        \hline\hline
%        $F_s$ (kHz) & 10 & 10 & 10 \\ % & 10 & 10 & 100 & 1000\\ 
%        \hline
%        $V_{walk}$ (m/s) & 1 & 5 & 10 \\ % & 0.5 & 0.1 & 1 & 1 \\
%        \hline
%        $f_d$ (Hz) & 16.7 & 83.3 & 166.7 \\ % & 8.3 & 1.7 & 16.7 & 16.7\\
%        \hline
%        $T_c$ (sec) & 0.025 & 0.005 & 0.003 \\ % & 0.051 & 0.25 & 0.025 & 0.025\\
%        \hline
%    \end{tabular}
%    \label{table:dataset}
%\end{table}

To better understand how CAA affects RF fingerprint authentication, 3 different CNN architectures were selected. A basic CNN consisting of two convolutional layers each with 64 neurons followed by a single dense layer forms the baseline performance comparison for the following models. The other models include VGG-16, a 16 layer neural network \cite{vgg16}; ResNet-50, a 50 layer network which introduces the concept of residual connections between layers \cite{resnet}; Inceptionv3, a deep CNN which utilizes a 'network within a network' strategy to learn features more deeply \cite{inception}; and Exception, an Inception based model which utilizes residual connections and separable convolutional layers to improve accuracy \cite{xception}. 
These models were originally intended for image classification, usually taking $244 \times 244 \times 3$ size inputs, hence modifications to the top layers were necessary to work with the received I/Q samples of size $1000 \times 8 \times 1$, where the 8 columns correspond to the I and Q signal samples from the 4 antenna elements. %The original Xception model architecture is shown in figure \ref{fig:xception}, where data first goes through the entry flow, then through a concatenation of 8 middle flows, then finally through the exit flow for classification. The final classifier for each network tested is a softmax layer consisting of 300 (the number of CAAs) neurons.

%\begin{figure*}
%    \centering
%    \includegraphics[width=0.8\textwidth]{xception.png}
%    \caption{Original Xception model architecture \cite{xception}. The model architecture first consists of the Entry flow, then Middle flow repeated 8 times, then finally Exit flow.}
%    \label{fig:xception}
%\end{figure*}

\begin{table}
\normalsize
    \centering
    \caption{Test Accuracy}
    \begin{tabular}
        {|c|| c |} %| c | c | c | c | c | c |} 
        \hline
        %\backslashbox{Name}{Set} & 1 & 2 & 3 \\ [0.5ex] %& 4 & 5 & 6 & 7 \\ [0.5ex] 
        Model & Accuracy \% \\
        \hline\hline
        CNN-3 & %79.4 \\ % & 93.4 & 98.2 & 66.4 & 43.2 & 42.2 & 
        93.3 \\
        \hline
        VGG-16 & %88.3 \\ % & 98.6 & 99.6 & 75.5 & 48.3 & 49.4 & 
        93.5\\
        \hline
        ResNet-50 & %92.3 \\ % & 99.6 & \textbf{100} & 78.2 & 57.7 & 57.2 & 
        99.2\\
        \hline
        Inception & %93.3 \\ %& 99.5 & \textbf{100} \\ % & \textbf{86.5} & 60.1 & 62.8 & 
        \textbf{99.9}\\
        \hline
        Xception & %\textbf{94.6} \\ %& \textbf{99.8} & \textbf{100} \\ % & 80.1 & \textbf{76.4} & \textbf{75.8} & 
        \textbf{99.9}\\
        \hline
    \end{tabular}
    \label{table:results}
\end{table}

\subsection{Results}

%Each network was trained to convergence, ranging from 10 iterations for the baseline CNN to 100 for Inception. Due to the depth of the VGG model, with approximately 138 million weighted parameters, it is impossible to train on such a small dataset in one attempt. To facilitate training the model, firstly, Everything except the top 7 convolutional layers removed and trained on the dataset. The weights from this training were transferred over to the full model, then frozen. With the top layers frozen, the the bottom of the network could be trained. The fully trained network was used for test classification performance. 
Table \ref{table:results} shows the test classification accuracy for each of the networks trained and tested in Python using TensorFlow. %on each dataset. 
Even the baseline CNN-3 scores significantly above the 63\% accuracy, which is the existing state-of-the-art performance in the literature achieved by ResNet-50 using the traditional (non-CAA) RF fingerprints \cite{9063411}. The %3 
more advanced networks, VGG-16 and ResNet-50, Inception, and Xception, all %have near perfect accuracy. %, with the only difference being datasets 5 and 6. It is not surprising that Xception scores highest out of 6 of the 7 datasets, as it combines architecture from both ResNet and Inception, allowing it to learn features more deeply than either one. The classification accuracy is extremely high, beating state-of-the-art non PUWS accuracy even on the hardest datasets.
achieve much higher accuracies. Comparing the performances of ResNet-50 using the CAA fingerprints ($>99\%$) to the traditional RF fingerprints ($63\%$) \cite{9063411} in similar setups, we see that the CAA fingerprints enable significantly enhanced authentication capacity with the help of deep neural networks.

%It is important to note that the classification accuracy depends heavily on $V_{walk}$ for our tests. With $F_s$ at 10kHz, it is counter-intuitive to think that a lower $V_{walk}$ results in lower accuracy. This is related to the channel coherence time $T_c$ being nearly equivalent to the sample duration  $T_{s} = N/F_s$ = 0.1sec. For $N$ = 1000 and $F_s$ = 10kHz, $T_c$ equals $T_{s}$ when $V_{walk}$ = 0.25m/s. Figure \ref{fig:ratio} shows how the accuracy for Xception varies with different ratios of $T_c$ to $T_{s}$. For fast fading scenarios ($T_c/T_s < 1$) the accuracy very quickly reaches 100\%, while for slow fading ($T_c/T_s > 1$), the accuracy, while not as sharp of an increase, reaches nearly 100\% at $T_c/T_s = 10$, when $V_{walk}$ = 0.05m/s. When $T_c$ = $T_s$, the samples in the authentication sequence become correlated, meaning that there is a statistically significant predictor function which relates samples in time. Depending on the magnitude of this correlation, the models will prefer to learn that correlation rather than the unique phase distortion from the RF fingerprints, leading to reduced performance. The further $|T_c/T_s|$ is from 1, the less correlated the samples are, leading to higher classification accuracy for all networks.

%\begin{figure}
%    \centering
%    \includegraphics[width=0.5\textwidth]{accuracy vs ratio.png}
%    \caption{Test set classification accuracy when sweeping $T_c/T_s$ from 0.005-25.4, ($V_{walk}$ = 0.01m/s to 50m/s). The minimum is 62.1\% accuracy at $T_c/T_s$ = 1}
%    \label{fig:ratio}
%\end{figure}

\section{Conclusion}
As dishonest actors develop more advanced security cracking technologies, it is vital that security keep in-step, lest vital systems become vulnerable to attack. More than ever there is a need for fast, cheap, secure authentication methods. In this vein, we investigated a novel authentication concept and implementation of RF fingerprinting through chaotic antenna arrays (CAAs). %We tested 5 different network architectures on 7 varying fast and slow fading scenarios. The results we have shown are promising for machine learning based automated authentication of these systems, with even simple networks performing extremely well. It is also seen that more advanced networks achieve perfect accuracy under a variety of scenarios. Performance under scenarios where channel coherence time nears sample duration could be improved; however, the results as they stand indicate that PUWS allows for very accurate RF fingerprint authentication in environments of up to 300 devices.
By testing 5 popular convolutional neural network (CNN) architectures, we showed that deep learning-based authentication utilizing CAA fingerprints significantly outperforms the existing state-of-the-art results using traditional RF fingerprints found in all communication devices. Compared to the 63\% accuracy \cite{9063411} achieved by ResNet-50, a popular CNN architecture, using the traditional RF fingerprints, the CAA fingerprints enable over 99\% accuracy by ResNet-50 in the task of authenticating 300 devices under Rayleigh fading channels. 

\printbibliography

\end{document}